# A self-rendering digital image encoding

Daniel L. Ruderman[*]
3765 Effingham Place
Los Angeles, CA 90027

*Abstract*

Without careful long-term preservation digital data may be lost to a number of factors, including physical media decay, lack of suitable decoding equipment, and the absence of software.  When raw data can be read but lack suitable annotations as to provenance, the ability to interpret them is more straightforward if they can be assessed through simple visual techniques. In this regard digital images are a special case since their data have a natural representation on two-dimensional media surfaces.  This paper presents a novel binary image pixel encoding that produces an approximate analog rendering of encoded images when the image bits are arranged spatially in an appropriate manner.  This simultaneous digital and analog representation acts to inseparably annotate bits as image data, which may contribute to the longevity of so-encoded images.

Keywords: digital preservation; rendering; image encoding; binary code

*1. Introduction*
The long-term preservation of digital data meets challenges on multiple fronts. First, the physical media on which we store data are volatile, with magnetic tapes and DVD-R's having lifetimes measured only in decades [1] [2].  Second, the equipment needed to read raw data from these types of media become quickly obsolete:  Locating hardware to read a 1980's era 8-inch floppy disk is not at present easy to do. Finally, even given the ability to read raw data it may not be obvious how to interpret them.  This last problem becomes compounded when centuries or millennia of time elapse and it may not be obvious that the media contain data at all. Labels on digital media will eventually disintegrate or fade and collections of microscopic data bits will then appear to the naked eye simply as attractive chromatic reflections.

The inability to directly interpret data in their intended manner is known as the *Viewing Problem* [3].  This might occur, for example, because the software used to format a word processing document or play an audio data file is no longer available.  However, as pointed out by Besser, some data representations offer extremely straightforward interpretation: "When we discover old film…, even if we don't have the right projector for that format, we can still hold it up to the light and see what's on it" [3]*.*  One recent solution to a Viewing Problem is the *Rosetta Disk* [4], which was created to preserve the interpretability of contemporary language into the distant future.  This nickel disk conspicuously displays a message in eight of the world's major languages as concentric spirals of ever-decreasing font size, leading eventually to microscopic texts in 1500 languages. At first glance this disk is clearly understood to convey hidden information at a smaller scale, thus compelling the observer to explore it further.  This paper addresses the Viewing Problem for digital images by leveraging this notion of "first glance interpretability" to devise a physical encoding of digital images which displays an analog representation when the bits are simply viewed by an observer.

---

[*] Current address: Center for Applied Molecular Medicine, Keck School of Medicine of USC, 2250 Alcazar Street, Los Angeles, CA 90033, ruderman@usc.edu.

### *2. Theory*

We describe a physical digital image encoding, called *contrast encoding*, that reveals to the unaided eye a semblance of the image whose pixels are coded digitally. Storage of digital image data typically employs hard disks, DVD-ROM's, or "flash" memory as media. But for long-term preservation we expect that longer-lasting media will be used. For example, durability over millennia can be achieved by etching image bits onto a durable surface (such as stone) or imprinting them in clay which is then fired. In this paper we assume that such a form of surface media is employed to store digital image data and that the *0* and *1* bits are visually contrasting to one another on the surface when viewed by eye. Contrast encoding has two required properties: (1) the spatial placement of image bits in their respective pixel positions, and (2) the use of a binary code which conveys the approximate pixel value when its rendered bits are spatially averaged.

The first property sets certain restrictions on the size of code words. For example, suppose an image is encoded as 9-bit grayscale. A pixel's 9 bits can be arrayed in a 3x3 grid to produce a square physical representation on the surface (assuming the bits themselves are square). The entire image can then be represented on the surface in its original aspect ratio by arraying these square pixel blocks in the same relative positions as the pixels they encode. Even when the number of bits per pixel is not a perfect square one can losslessly multiply the pixel values by an appropriate power of two to achieve the next pixel bit depth that is a perfect square.

The second property of contrast encoding is meant to provide surface reflectances which spatially average to an approximate representation of the pixels' values. To achieve this, we note that binary code words for pixel values admit a symmetry through their permutation; we are free to choose which code word represents which pixel value. The integer binary format used in digital computers (typically two's complement) is convenient for performing numerical operations. However, to better render pixel values it will be beneficial to employ a binary code that has high correlation between the number of *1*'s in a code word and the pixel value it represents. The binary code presented here is one in which there is a non-decreasing relationship between the number of *1*'s in the binary word and its value. This property enables a contrast in surface reflectance between *0*'s and *1*'s to render an analog approximation to the digitally encoded pixel values. The fidelity of this analog representation is limited by quantization to the *b+1* different levels available in the code word reflectances.

The contrast encoding work flow is shown in Figure 1. Each pixel in an image has the code word corresponding to its value looked up in a dictionary with a one-to-one map between pixel values and code words. The square code word is rendered on the surface of long-lasting media, such as stone, using a method that visually contrasts between *0* and *1* bits. Once rendered in this way, the image will be visible to the eye in approximate form when the surface is viewed in plain lighting.

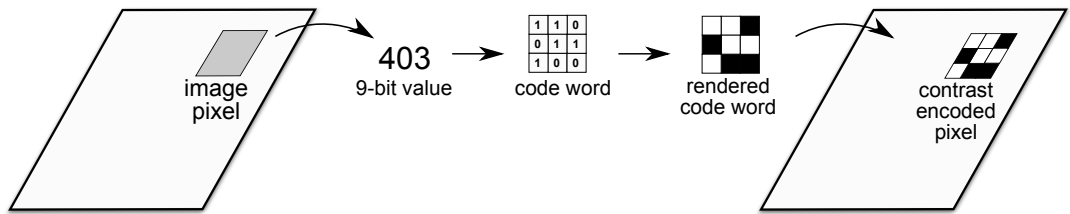

*Figure 1:* Contrast encoding work flow. The value of an image pixel (403) is used to look up a code word (110011100) from a pre-defined dictionary. The binary code word is then converted to a rendered code word by assigning *1*'s to a high reflectance and *0*'s to a low reflectance. The square code word is then rendered on the surface of long-lived media in the same spatial location as the original pixel. This process is repeated for all pixels in the image.

To be precise, let *b* be the number of bits per word (bit depth) used to encode pixels whose values range from *0* to $2^b-1$. There are $bCn$ ways to distribute exactly *n* *1*'s among the *b* bits. So there is one word which is all *0*'s, *b* words containing one *1*, and *b(b-1)/2* words containing two *1*'s. To achieve high correlation between the average surface reflectance of the *b* bits in a word with the value it encodes we assign words with fewer *1*'s to lower values (assuming that *1* bits reflect more strongly than *0* bits). We use a code in which the pixel value *0* is represented by the word which is all *0*'s, the *b* values ranging from *1* to *b* by the *b* words containing one *1*, the *b(b-1)/2* values from *b+1* to *b(b+1)/2* by words containing two *1*'s, and so on, with the highest pixel value $2^b-1$ being represented by the word containing *b* *1*'s.

An encoding for the case *b=4* is shown in Figure 2. In this example the contrast encoded values have a non-decreasing count of *1*'s as a function of pixel value. The difference between the standard binary encoding and contrast encoding methods is apparent in Figure 3, which shows the relationship between pixel value and the number of *1*'s in an encoding of 9-bit pixels. While the *1* count in the ordinary encoding is weakly correlated with pixel value (Pearson correlation r=0.58), it is not as strong as that in the contrast encoding (r=0.96). Furthermore, the standard encoding has many reversals in *1* count as a function of pixel value, making it inappropriate for visually encoding pixel value. Thus image contrast will be larger when viewing the encoded surface using contrast encoding over that in the standard one. While this example is specific to grayscale images the same method can be generalized to color, as will be discussed in Section 3. A similar effect could be achieved by representing the image by a unary encoding, in which the number of *1*'s used to represent a pixel is exactly the pixel's value. This would have the advantage of an exact image rendering in $2^b$ gray levels rather than the *b+1* levels for the contrast encoding. However, this would come with the associated cost of an exponentially larger number of bits per pixel, since $2^b$ bits would be needed per pixel to represent all the gray levels. A hybrid method that interpolates between these two coding extremes may offer an advantageous compromise. Note that in contrast encoding there is remaining arbitrariness in the value-to-word mapping because all of the words with the same number of *1*'s belong to an equivalence class with respect to rendering by reflectance and therefore will achieve the same visual effect if their pixel values are permuted. A randomized pixel value assignment to code words may be preferred, for example, because it reduces rendering artifacts like rasterization.

For *b>>1* the values of $bCn$ are sharply peaked at *n*'s which are within a value difference of order $b^{1/2}$ away from *b/2*, so most code words will render contrasts within this limited range. Thus a typical contrast encoded image will have a visual contrast of order $b^{-1/2}$ relative to its mean. This will tend to create a loss of contrast in the rendered image relative to the original image, particularly for large *b*. Therefore a trade-off exists between a highly accurate digital representation using a large bit depth and a high contrast analog representation using a smaller bit depth (though with an associated increase in quantization artifacts). Since the bit depth of digital images is typically low (in the 8- to 16-bit range), the non-uniformity of available code word contrasts for very large *b* is not likely to be a highly restrictive property of contrast encoding. Note that the visual contrast is fundamentally limited by the apparent contrast between *0* and *1* bits as rendered in the medium.

| Pixel value | 0 | 1 | 2 | 3 | 4 | 5 | 6 | 7 | 8 | 9 | 10 | 11 | 12 | 13 | 14 | 15 |
|---|---|---|---|---|---|---|---|---|---|---|---|---|---|---|---|---|
| Contrast encoding | 0000 | 0001 | 0010 | 0100 | 1000 | 0011 | 0110 | 0101 | 1001 | 1010 | 1100 | 0111 | 1011 | 1101 | 1110 | 1111 |
| *1* count | 0 | 1 | 1 | 1 | 1 | 2 | 2 | 2 | 2 | 2 | 2 | 3 | 3 | 3 | 3 | 4 |
| Spatial representation | | | | | | | | | | | | | | | | |

*Figure 2*: Example contrast encoding for *b=4* bits in a 2x2 bit spatial representation with black and white pixels corresponding to *0* and 1 bits, respectively.

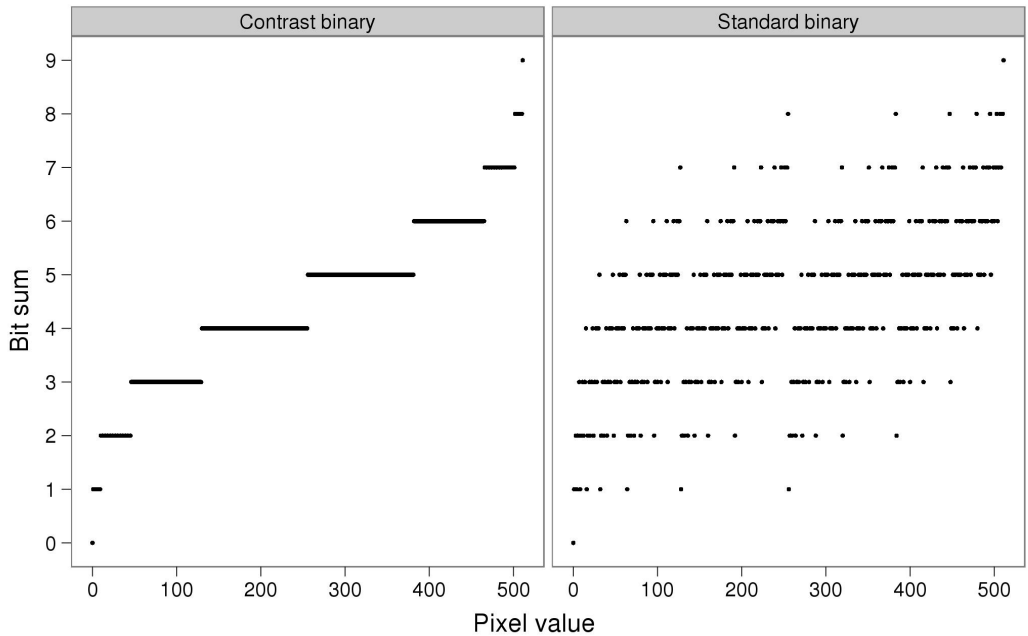

*Figure 3:* Bit sum of pixel encoding words. Contrast encoding (left, Pearson correlation r=0.96) and standard encoding (right, r=0.58) for 9-bit pixel values ranging from 0 to 511.

### *3. Results*
Figure 4A shows part of the Mona Lisa, in 9-bit grayscale (top) together with the same region encoded using a 3x3 (9-bit) contrast encoded binary representation, shown blurred and subsampled to the original resolution (bottom). Blurring was performed using a Gaussian filter with FWHM=5 pixels, which was chosen to be somewhat larger than the size of a code block to simulate distance viewing and to prevent aliasing during subsampling. From this example it is evident that the contrast encoded image presents a semblance of the image when viewed by eye from a distance suitable to visually merge individual bits. Figure 4B shows an expanded view of Mona Lisa's smile, providing a view of the individual bits that constitute the encoded image. When viewed at varying distances one can discern the transition from bit viewing to the bit averaging which yields the analog rendering. Figure 4C shows a closer view of 3x3 pixels (top) together with the 9 corresponding code words, each in a 3x3 binary representation. Note that the darkest pixel in the image (lower left) has three white pixels in the encoding, while the others have four. The relatively coarse contrast scale present in the bit image is exemplified by the fact that these eight pixels each have different values but all are displayed with four white and five black pixels.

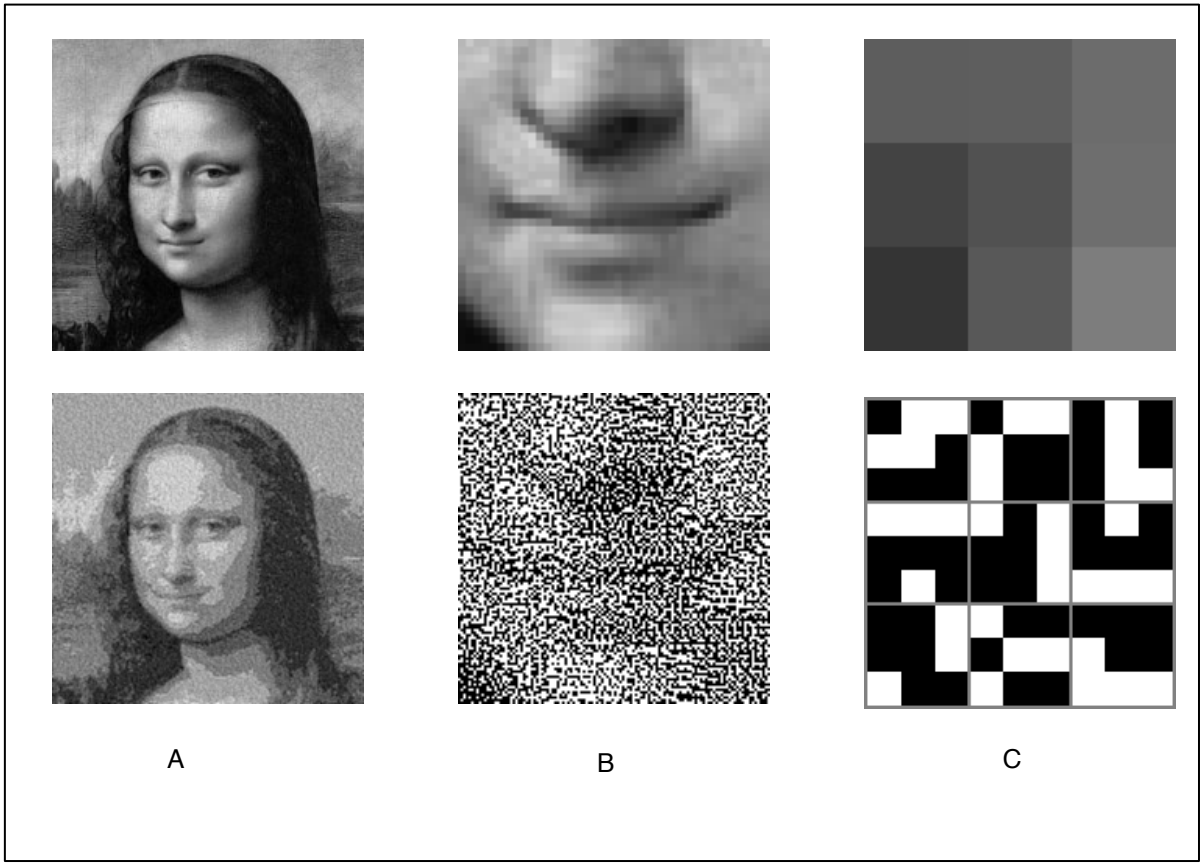

*Figure 4:* Contrast encoding of Mona Lisa. A: 200x200 pixel section of painting in grayscale (top); contrast encoded image, blurred using a FWHM=5 Gaussian filter and subsampled from full 600x600 encoded format to match original image size (bottom). B: 100x100 pixel subsection of the smile, up-sampled by factor of 3 to match encoded resolution (top); contrast encoding of smile, at full 300x300 resolution (bottom). When viewed at a distance the similarities between the top and bottom images become more pronounced. C: 3x3 pixel detail at right mouth corner (top); 9x9 encoded image of smile corner, showing binary code for each of the 9 pixels, with gray lines delineating individual 3x3 pixel encodings (bottom). Encoded *1* bits are displayed as white and *0* bits as black.

It is straightforward to extend this method to color images. Their physical embodiment would require a medium that can display at least four different values, one black and the other three as color primaries. Colored tiles, for example, would be able to achieve this effect. Such an encoding would need all three color channels for each pixel to be placed in close spatial proximity. This can be achieved if each pixel is represented with a number of bits that is both a multiple of three and the square of an integer. While nine total bits would allow for an unacceptably low three bits per color channel, a total of 36 bits would increase this to 12, which is sufficient for most images. An example of this 36-bit encoding is shown in Figure 5. The original 8 bit per color channel image was first converted to a 12 bit representation by multiplying each value by 4. The R, G, and B channel values for each pixel were placed together in a 6x6 bit array to represent the pixel. The 12 bits for each color channel occupy two adjacent 6-bit columns, and the ordering of these column pairs was chosen randomly for each pixel to avoid vertical raster artifacts caused by each image column containing a unique color. The original image and its encoded version (blurred using a Gaussian filter with FWHM=9, subsampled to the original 300x300 pixel resolution, and multiplicatively brightened) are shown in Figure 5A top and bottom, respectively. The loss of contrast in the encoded image is apparent. In Figure 5B is shown an expanded view of the 9 pixels in the center of the star, both in the original (top) and encoded (bottom) formats. As in the case of the grayscale image, this method of representing a color image provides both a clear indication of the contents of the encoded image (albeit with lower contrast) and preserves the image data in the details of the encoding.

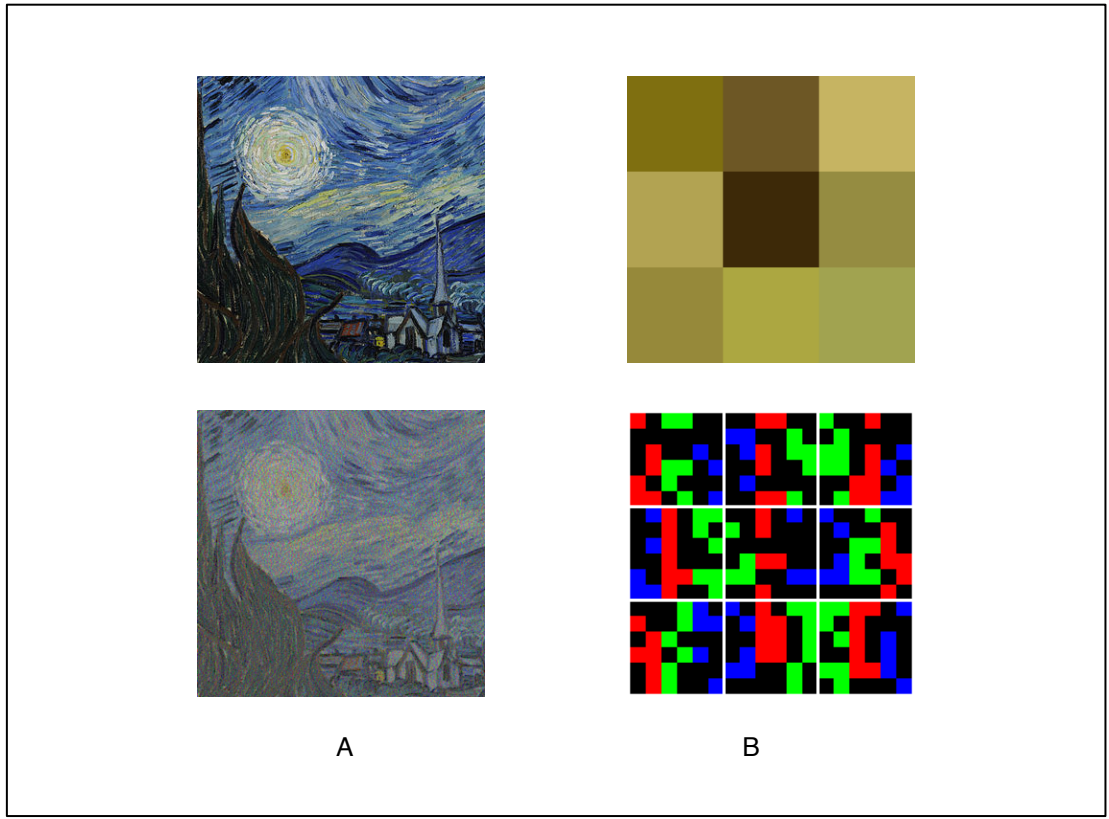

A                    B

*Figure 5:* Example of color image encoding. A: Detail from van Gogh's *Starry Night*, original at 300x300 pixels (top) and contrast encoded at b=12 bits per color channel, blurred using a FWHM=9 Gaussian filter, and down-sampled from 1800x1800 to match original (bottom); encoded image pixel values have been multiplied by 3 to correct for intensity loss of encoding. B: Original (top) and encoded (bottom) 3x3 pixel array from center of star in A; 6x6 binary pixel encodings are separated by white lines.

To completely decode the data the observer must be able to assign a pixel value to each pattern of bits comprising an encoded pixel. While it is readily apparent that brighter encoded pixels (i.e. more *1's* in the encoding) should be assigned higher values than the dark ones, there is no explicit information about the relationship between each code word and its pixel value. One way the encoding could be inferred is to apply prior knowledge that contrast gradients in typical images tend to vary smoothly across many pixels; this observation would allow the observer to order the code words by value through statistical analysis. However, a more basic solution would be to include alongside the image a separate pixel value "key", for example by rendering all code words in a linear array from darkest to brightest. A key such as this would be fairly straightforward to recognize and interpret.

## *4. Discussion*

This paper presents a method for the binary encoding of images that aids their long-term preservation and interpretation. The method employs a data format whose bit representation is readily interpreted as an encoded image when made visible to the unaided eye. Thus the encoding is simultaneously digital and analog. The only requirement for realizing this encoding is the use of a physical encoding medium that can display visually contrasting and long-lived bit patterns, for example etched stone. Images synthesized by the encoding have a similar appearance to half-toned pictures or ASCII art, though with the additional feature that the image is preserved in a binary encoding without added per-pixel redundancy. Since the physical representation of the encoded image mimics the image itself this representation can be considered a visual form of onomatopoeia.

Like the Rosetta Disk, physical representations of contrast encoded images create an immediate impression of the information contained therein. But unlike the inward text spirals on the Disk, it is somewhat less obvious that further exploration of contrast encoding images will provide additional information. It would be left to the viewer's curiosity to discover the lossless digital encoding of the image, thus "bootstrapping" the decoding [5]. Even in the absence of a complete decoding, the viewer can glean a great deal about a contrast encoded image simply by looking at its bits. This analog image rendering is an intrinsic annotation of the digital image data. However, unlike a paper label which will one day crumble or fade, this form of annotation will last exactly as long as the bits do, and is expected to degrade gracefully over time along with the bits themselves.

## *Acknowledgements*

I thank K. D. Bollacker, M. R. DeWeese, A. Golomb, A. Pelah, D. L. Ringach, and L. J. Williams for helpful comments on the manuscript.